\documentclass{article}
\usepackage{spconf,amsmath,epsfig,subfigure,amssymb}
\usepackage[latin1]{inputenc}

\renewcommand{\vec}[1]{{\mathbf {#1}}}

\title{Trellis-Coded Quantization for Public-Key Steganography}
\name{Ga\"etan Le Guelvouit}
\address{
	{\tt leguelvouit@lss.supelec.fr}\\
	Laboratoire des Signaux et Syst\`emes -- Sup\'elec\\
	3, rue Joliot-Curie -- 91192 Gif-sur-Yvette -- FRANCE
}

\begin{document}
	\maketitle

\begin{abstract}
This paper deals with public-key steganography in the presence of a passive warden. The aim is to hide secret messages within cover-documents without making the warden suspicious, and without any preliminar secret key sharing. Whereas a practical attempt has been already done to provide a solution to this problem, it suffers of poor flexibility (since embedding and decoding steps highly depend on cover-signals statistics) and of little capacity compared to recent data hiding techniques. Using the same framework, this paper explores the use of trellis-coded quantization techniques (TCQ and turbo TCQ) to design a more efficient public-key scheme. Experiments on audio signals show great improvements considering Cachin's security criterion.
\end{abstract}
\section{Introduction}\label{sec:intro}
Steganography is the art of hiding secret information within innocuous documents. It is often schemed by Simmons' prisoners' problem~\cite{Simmons83}. Alice and Bob are in prison and want to finalize a common escape plan, but their communications are filtered by a warden named Wendy. If she considers a transmitted document suspicious, she stops the communication channel. Prisoners must exchange secret information using innocuous contents. This purpose is the definition of steganography. Unlike watermarking, steganography does not involve any robustness. Its goal is transparency: statistics of stego-signals must look as natural as original cover-signals, and visual (or auditive) quality can not suffer of any default.

Whereas a lot of solutions for symmetric steganography (i.e. a private key is shared by Alice and Bob) have been proposed, very few practical proposals for the asymmetric version (a public key for embedding and a secret key for reading) exist. The problem has been theoretically analyzed, but -- as far as we know -- only one practical attempt has been done~\cite{Guillon02}. However that work suffers of important limitations. In particular, its transparency depends of a important knowledge of cover-signal statistics. 

The aim of this article is to design a general purpose public-key scheme. Section~\ref{sec:si} recalls the problem model, i.e. communication through channel with side information. Next section presents a practical solution to the public-key steganography problem. Section~\ref{sec:conclusion} concludes this work.

\section{PROBLEM MODEL: CHANNEL WITH SI}\label{sec:si}
Alice wants to transmit a message $\vec m$. To this end, it is first encoded to $\vec w \in \mathbb R^n$. Let us consider an i.i.d. cover-signal denoted $\vec x \in \mathbb R^n$, modeled by the random variable $X$. Whereas steganographic schemes are not supposed to be robust (i.e. resilient to attacks like lossy compression, filtering, etc.), our initial assumption on host signal statistics implies the use of a de\-correlating transform, resulting an additional quantization noise denoted $\vec z$, which is modeled by $Z \sim \mathcal N(0, N)$. The resulting stego-signal is denoted $\vec y' = \vec w + \vec x + \vec z$. In classical encoding scheme, the capacity of such channel is very low due to invisibility constraint. Nevertheless, since $\vec x$ is perfectly known at the encoder, steganography represents a channel with side information available at the encoder. As demonstrated by~\cite{Costa83,Cohen02DP}, its capacity is
\begin{equation}
	C = \frac 1 2 \log_2 \left[
		1 + \frac P N
	\right] \textrm , \label{eq:capa}
\end{equation}
where $P$ is the embedding power constraint $\sum_{i=1}^n \vec w[i]^2 \leq P$. Therefore, side information $\vec x$ does not have any influence on the capacity. The use of informed encoding may then lead to important capacity or invisibility improvements for steganography.

Practical -- but sub-optimal -- schemes for side informed encoding have then been proposed by watermarking community. A famous one is scalar Costa scheme (SCS)~\cite{Eggers02}. Its principle is to use scalar quantization to define an informed codebook. For simplicity, let us consider binary transmission: $\vec m \in \{ 0 , 1 \}^n$. For a given quantization step $\Delta$, SCS defines $\mathcal U$, product of scalar codebooks $\mathcal U = \mathcal U[1] \times \mathcal U[2] \times \ldots \times \mathcal U[n]$ with 
\begin{equation}
	\mathcal U[i] = \left\{
		k \frac \Delta 2 + \mathbf d[i] \textrm{,~} k \in  \mathbb Z
	\right\} \textrm , \label{eq:scs_dither}
\end{equation}
where $\vec d \in [- \Delta/2 , + \Delta/2]^n$ is dither noise used as a private key. Each possible message $\vec m$ is associated to a sub-codebook $\mathcal U_{\vec m} \subset \mathcal U$, defined by 
\begin{equation}
	\mathcal U_{\mathbf m}[i] = \left\{
		k \Delta + \mathbf d[i] + \frac{\Delta \mathbf m[i]} 2 \textrm{,~} k \in  \mathbb Z
	\right\} \textrm .
\end{equation}
To encode $\vec m$, the chosen codeword is defined as 
\begin{equation}
	\vec u^{\star} = \arg \min_{\vec u \in \mathcal U_{\vec m}} \| \vec u - \vec x \|^2 \textrm , 
\end{equation}
and the added signal is $\vec w = \alpha (\vec u^{\star} - \vec x)$.

Experiments show that SCS poorly performs for uncoded messages: for an embedding rate of 1~bit per cover-element, $P/N$ must be greater than 14~dB to get a bit error rate lower than $10^{-5}$ ($9.2$~dB away from theoretical capacity). It must be associated to a efficient channel code, but this reduces embedding rate. 

\section{PUBLIC-KEY STEGANOGRAPHY}\label{sec:wt}
Symmetric steganography suffers of an important drawback: Alice and Bob must share a secret key before any secret transmission. Since their communication are supervised by Wendy, this secret must be transmitted before their imprisonment. Public-key steganography permits to avoid this transmission. Each transmitter owns a pair of cryptographic keys $(\vec k_{\scriptsize \textrm{pub}}, \vec k_{\scriptsize \textrm{prv}})$. The public one can be freely communicated without any impact on secrecy, and $\vec k_{\scriptsize \textrm{prv}}$ must be kept secret. Everyone can send a secret message to Bob, but only Bob is able to read it.  
        
	\subsection{Previous work}
\begin{figure}
		\begin{center}
			\includegraphics[width=8cm]{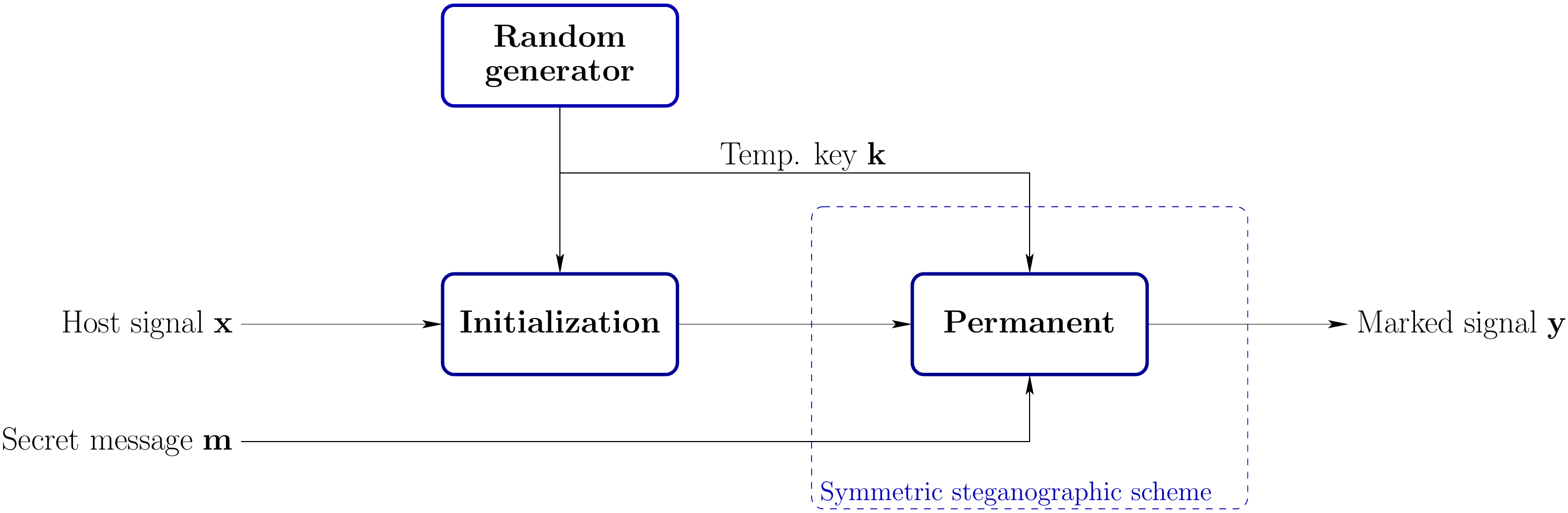}
			\caption{Public-key framework from~\cite{Guillon02}. Permanent phase is initialized
			by a random temporary key $\vec k$.}
			\label{fig:asym}
		\end{center}
	\end{figure}

From ideas of \cite{Anderson98}, Guillon {\it et al.}~\cite{Guillon02} proposed a practical framework for public-key steganography, based on asymmetric cryptography and a steganographic scheme using SCS. It consists in two steps (see Fig.~\ref{fig:asym}): 
\begin{description}
	\item[Initialization phase] A secret key $\vec k_{\scriptsize \textrm{tmp}}$ is randomly chosen. It is encrypted using an asymmetric crypto-algorithm with public key $\vec k_{\scriptsize \textrm{pub}}$. The random-like binary vector $\vec k' = \textrm{crypt}(\vec k_{\scriptsize \textrm{tmp}}, \vec k_{\scriptsize \textrm{pub}})$ is embedded into cover-signal. 
	\item[Permanent phase] The secret message $\vec m$ is transmitted using SCS. Embedding is done by using a secret dither noise $\vec d$ (see Eqn.~(\ref{eq:scs_dither})) which is generated with the seed $\vec k_{\scriptsize \textrm{tmp}}$.
\end{description}
Whereas the second step do not represent any major issue (SCS with secret dithering presents good security properties~\cite{Guillon02}), initialization phase needs to embed public information without any noticeable change on stego-signal statistics and quality. For that purpose, Guillon {\it et al.} considered the use of SCS embedding with $\alpha = 1/2$. But in the case of non-uniform cover-signal pdf, this leaves easily noticeable artifacts on stego-signal statistics (e.g. for Gaussian case, see Fig.~\ref{fig:pdf:a}). This problem does not appear in the permanent phase thanks to the secret dither noise $\vec d$. Their solution is to use a compressor before embedding to equalize the pdf, then embed $\vec k'$ and apply the inverse compressor to get the original pdf shape. This implies that compressor design must be robust to data embedding and quantization noise. 

	\subsection{Trellis-coded quantization for initialization phase}
\begin{figure}[t]
		\centering
		\subfigure[SCS embedding with $\alpha = 0.5$]
		{
			\includegraphics[width=6.5cm]{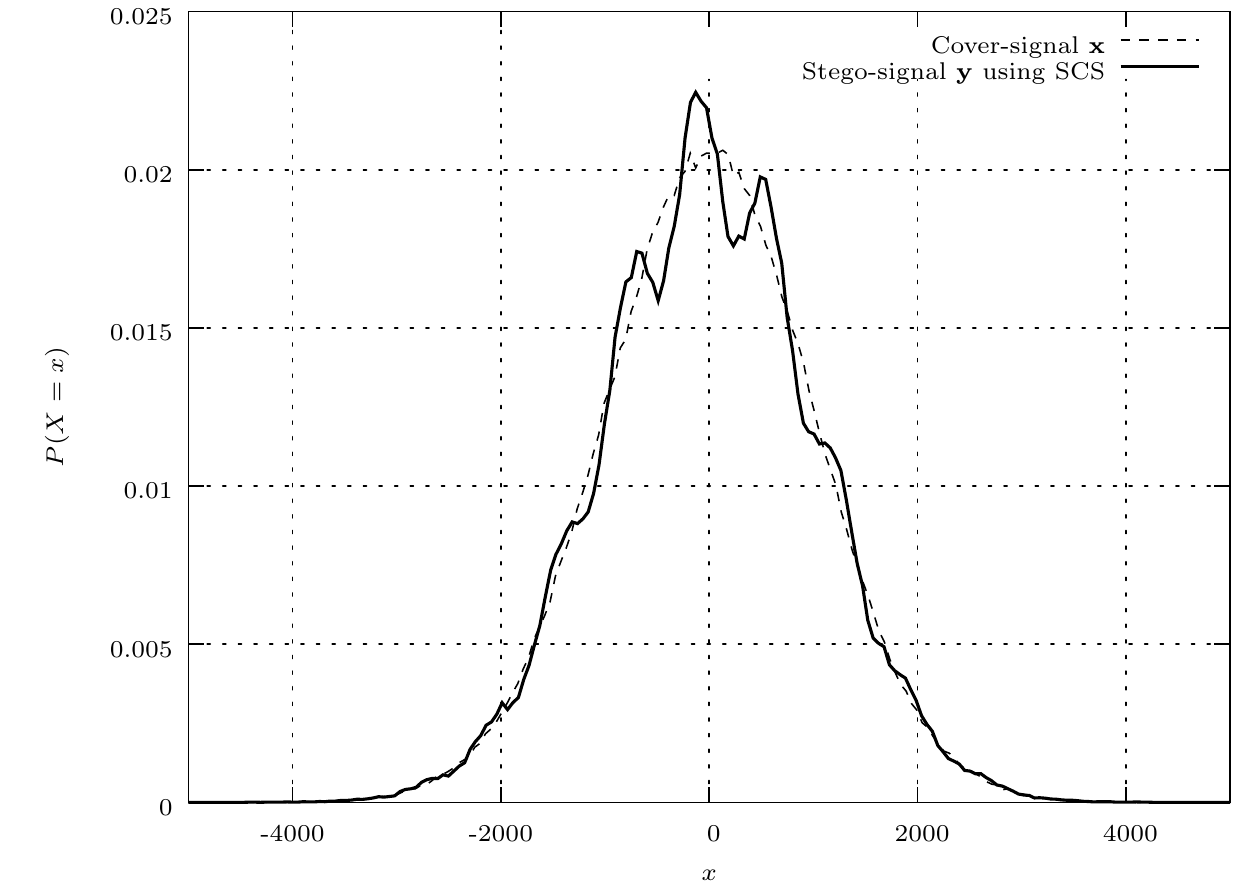}\label{fig:pdf:a}
		}
		\subfigure[Embedding using TCQ codes with $\alpha=0.7$]
		{
			\includegraphics[width=6.5cm]{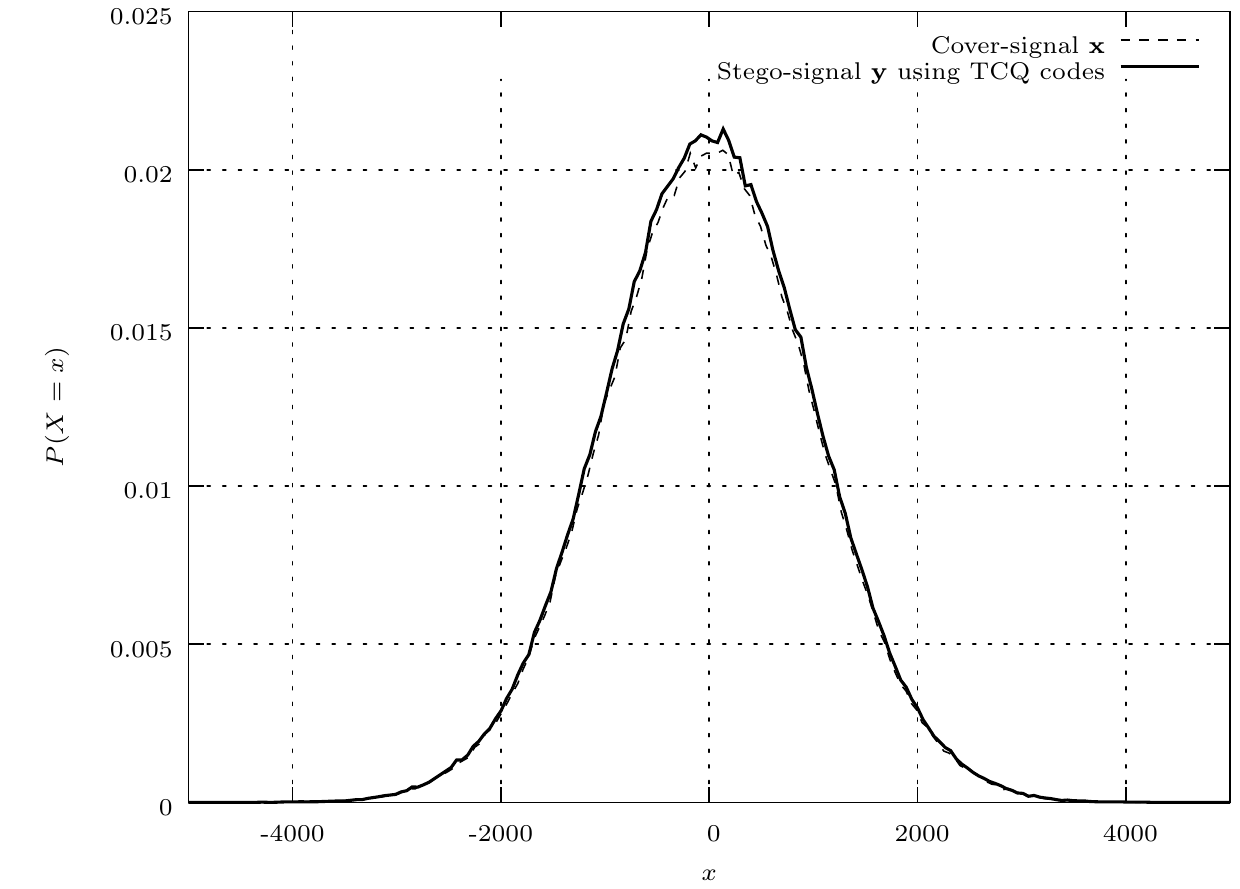}\label{fig:pdf:b}
		}
		\caption{Resulting probability density functions after 
		embedding ($X \sim \mathcal N(0, 10^6)$ and $P=10^4$).
		A $2^{9}$-state trellis is used for TCQ.}
		\label{fig:pdf}
	\end{figure}

Since SCS partioning is regular, it introduces artifacts on marked signals. The approach that we propose is to use trellis-coded quantization to design a pseudo-random space partitioning. TCQ techniques for side informed coding combine robustness, SCS-like capacity and ease of implementation. Let us consider a trellis defined by a transition function: 
\begin{eqnarray}
	\mathcal S \times \{ 0, 1 \} & \longrightarrow & \mathcal S 
	\nonumber \\ 
	t : \left(     
		s_i, \vec m[i]
	\right) & \longmapsto & s_{i+1} \textrm ,
\end{eqnarray}
where $\mathcal S = \{ 0, 1, \ldots , 2^{r-1} \}$ is the set of possible trellis states. Unlike SCS, dither noise $\vec d$ is no more random but a function of current state and input symbol (a private key may be also introduced in the function for additional security purpose):
\begin{eqnarray}
	\mathcal S \times \{ 0, 1 \} & \longrightarrow &  [- \Delta/2 , + \Delta/2]
	\nonumber \\ 
	o : \left( 
		s_i, \vec m[i]
	\right) & \longmapsto & \mathbf d[i] \textrm .   
\end{eqnarray}
Sub-codebooks are then defined by $$\mathcal U_{\mathbf m}[i] = \left\{ k \Delta + o(s_i, \mathbf m[i]) \textrm{,~} k \in  \mathbb Z \right\}$$ and closest codeword $\vec u^{\star} \in \mathcal U_{\mathbf m}$ to $\vec s$ is computed using a Viterbi algorithm with strong {\it a priori} metrics to ensure decoded codeword to belong to $\mathcal U_{\vec m}$: 
\begin{equation}
	\vec u^{\star} = \arg \min_{\vec u \in \mathcal U_{\mathbf m}} \sum_{i=1}^n \left( 
		\vec s[i] - \vec u[i]
	\right)^2 \textrm . \label{eq:ustar}
\end{equation}
Stego-signal is given by $\vec x = \alpha \left( \vec u^{\star} - \vec s \right)$. Experiments show that the best $\alpha$ parameter in term of robustness is $P/(P+N)$  like in the original Costa scheme~\cite{Costa83}. As demonstrated by Fig.~\ref{fig:pdf:b}, no statistical artifact is noticeable using this technique. 

On the other hand, the use of linear embedding (i.e. $\vec w = \alpha (\vec u^{\star} - \vec x)$) leaves another clue for Wendy. Let $d_{\vec x}$ be the distance between an original cover-signal $\vec x$ and its closest codeword $\vec u \in \mathcal U$. We have $\mathbb E\left[ d_{\vec x}\right] = \Delta^2/48$ (since two quantizers are available for each input data, thanks to trellis coding). Let $d_{\vec y}$ be the distance between a stego-content $\vec y$ and its closest codeword. To avoid suspicion, we must ensure $d_{\vec y} \simeq d_{\vec x}$, i.e. use $\alpha=1/2$. But this parameter does not permit to reach the Voronoi region associated to $\vec u^{\star}$ in most practical cases. A larger $\alpha$ parameter must be chosen, leading to $d_{\vec y} < d_{\vec x}$.

\begin{figure}
		\begin{center}
			\includegraphics[width=6cm]{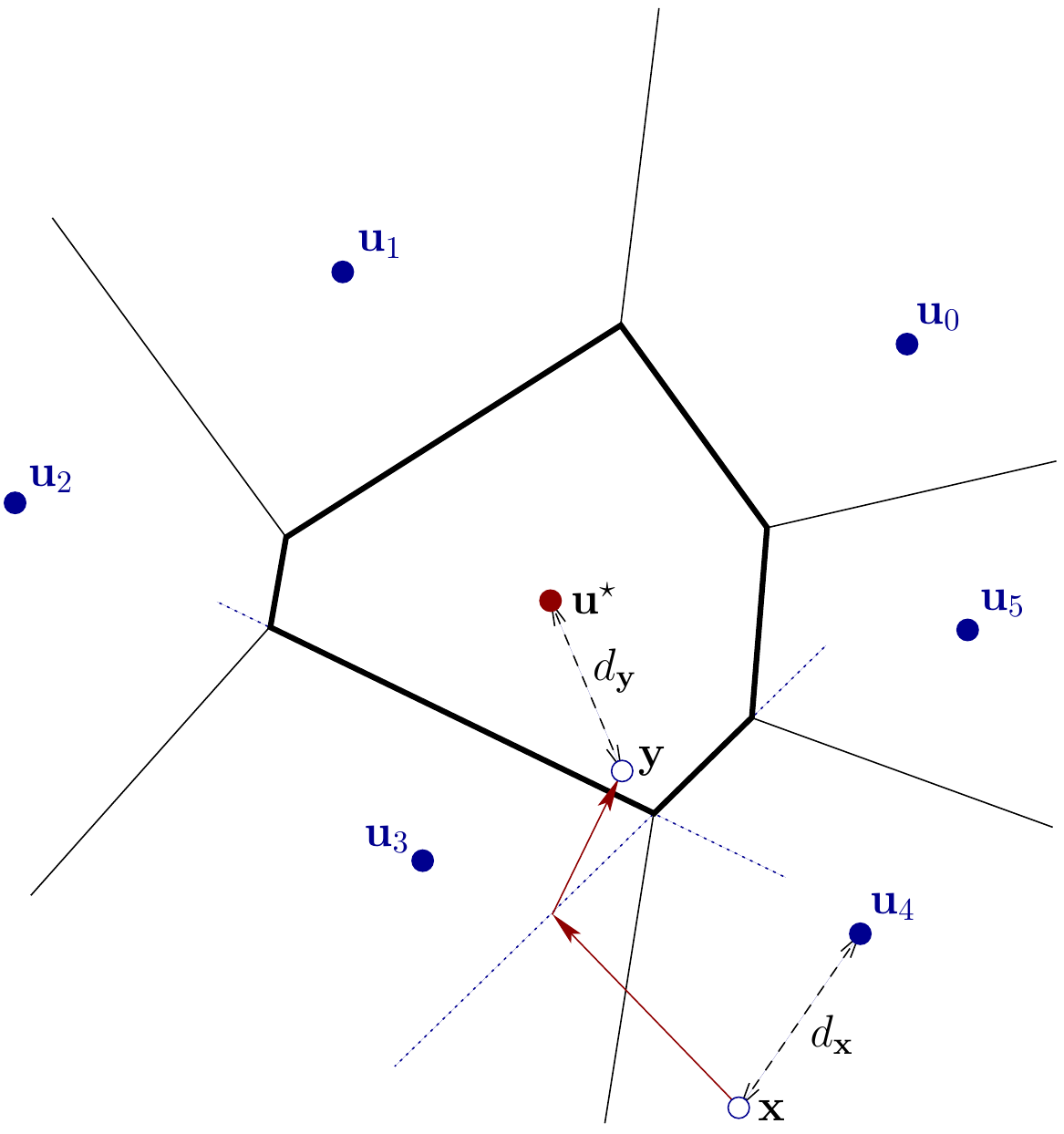}
			\caption{Data embedding for initialization phase using a Monte Carlo technique.}
			\label{fig:mc}
		\end{center}
	\end{figure}

The chosen embedding technique is similar to Miller's iterative solution~\cite{Miller04} and it is illustrated by Fig.~\ref{fig:mc}. The idea is to bring cover-signal $\vec x$ into the chosen area, at a distance $\sqrt N + \epsilon$ of its frontier\footnote{Since the realization of $Z$ is unknown during embedding, we introduce $\epsilon$ to have an additional security margin.}. It iterates as follows:
\begin{enumerate}
	\item Let $\vec y = \vec x$ and let $\vec u^{\star} \in \mathcal U_{\vec m}$ be the targeted codeword.
	\item Find the closest codeword $\vec u \in \mathcal U$ to $\vec y$. If $\vec u = \vec u^{\star}$, stop.
	\item Let $\displaystyle \vec d = \frac{\vec u^{\star} - \vec u}{|\vec u^{\star} - \vec u|}$ and $\displaystyle \beta = \frac{|\vec u^{\star} - \vec u|}{2} + \sqrt N + \epsilon$.
	\item Add $\beta \vec d$ to $\vec y$ and go to step 2.
\end{enumerate}
Obviously, this technique is not very efficient in term of capacity. Nevertheless, the size of the message transmitted during this first phase is not large, since it just represents a pseudo-random generator seed. 


	\subsection{Powerful dirty paper codes for permanent phase}
Turbo TCQ~\cite{Chapel03} is a recent source coding technique inspired by iterative channel decoding algorithms. The turbo trellis-coded quantizer is composed of two parallel TCQ trellises. The first one works with the signal $\vec y'$ to be decoded, while the second one decodes an interleaved version of $\vec y'$. {\it A posteriori} metrics from first quantizer are used as {\it a priori} metrics for the second. The process is repeated until both {\it a posteriori} metrics are similar. 

\begin{figure}[t]
		\begin{center}
			\includegraphics[width=7cm]{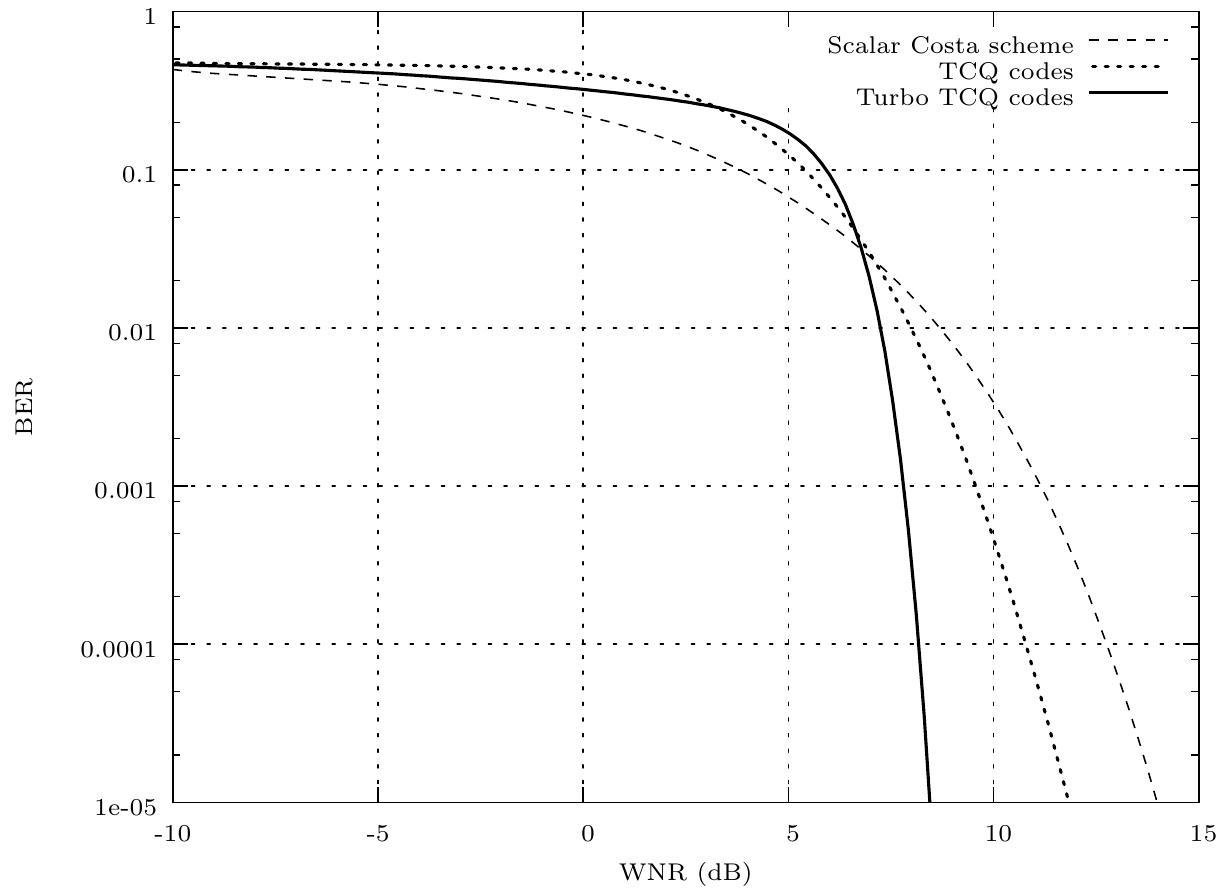}
			\caption{Bit error rates for trellis-coded quantization codes compared to SCS. Embedding rate is 1~bit per cover element.}
			\label{fig:peb}
		\end{center}
	\end{figure}

We use this technique to design a powerful dirty paper code. For an embedding rate of 1~bit per cover element, the use of turbo TCQ leads to a gain of $5.5$~dB compared to classical SCS (see Fig.~\ref{fig:peb}). Since robustness is not a priority in the case of steganography, this gain leads to a better transparency.  

	\subsection{Results: security from Cachin's point of view}
\begin{figure}[t]
		\begin{center}
			\includegraphics[width=7cm]{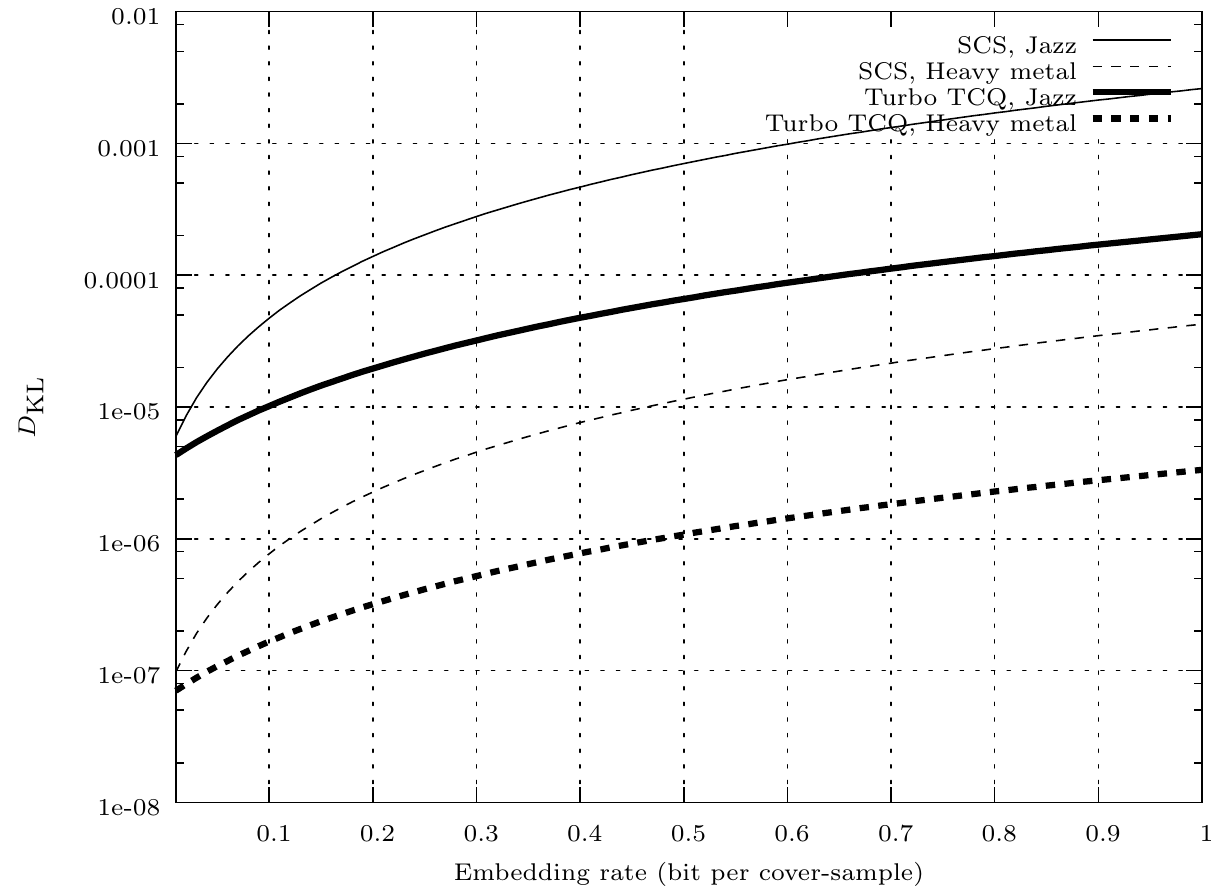}
			\caption{Performance of permanent phase in term of $\epsilon$-security for SCS and proposed technique.}
			\label{fig:kl_perm}
		\end{center}
	\end{figure}

Cachin's paper~\cite{Cachin98} is a pioneer try for the definition of a security criterion for steganography. Considering i.i.d. cover-signal (like this work), it defines security as the relative entropy between cover and stego-signal. A steganographic scheme is said to be $\epsilon$-secured against passive warden if $D_{\scriptsize \textrm{KL}}(P_Y \| P_X) \leq \epsilon$, where 
\begin{equation}
	D_{\scriptsize \textrm{KL}}(P_Y \| P_X) = \sum_{c \in \mathcal C} P_X(c) \log \frac{
		P_X(c)
	}{
		P_Y(c)
	}\textrm .
\end{equation}
In order to evaluate $\epsilon$-security of the proposed permanent phase, we used our technique for audio steganography. Two test samples were used: one is a smooth bass solo ("Jazz") and the other is an powerful guitar play ("Heavy metal"). Both are 5 seconds long PCM samples, 44.1 KHz -- 16 bits. Embedding is performed like Guillon's practical proposition: MDCT on analysis windows of 512 samples, and coefficients are grouped into 32 sub-bands during 10 windows (i.e. each sub-band contains 160 coefficients). Each sub-bands are supposed to be Laplacian distributed.
A random binary message is embedded using SCS and turbo TCQ codes. Quantization step $\Delta$ is chosen to get a bit error rate lower than $10^{-5}$, and embedding rate is modified using spread transform (like ST-SCS~\cite{Eggers02}). Fig.~\ref{fig:kl_perm} shows the resulting $\epsilon$-security. We can see that the security gain is similar for both test samples. From Cachin's point of view, our turbo TCQ codes are about ten times much secure than classical SCS for high embedding rates.

\section{CONCLUSION}\label{sec:conclusion}
This paper provided a practical and efficient solution to the public-key steganography problem. Within an initial framework based on asymmetric cryptography, we improved its generality and its efficiency. Unlike~\cite{Guillon02}, the proposed initialization phase (transmission of the secret key) is independent of cover-signal statistics thanks to a pseudo-random space partitioning. And the permanent phase (transmission of the secret message) is more secure concerning Cachin's criterion.  

	\bibliographystyle{IEEEbib} \small
	\bibliography{main}

\begin{thebibliography}{1}

\bibitem{Simmons83}
G.~J. Simmons,
\newblock ``The prisonners' problem and the subliminal channel,''
\newblock in {\em Advances in Cryptology: Proc. of CRYPTO}, 1984, pp. 51--67.

\bibitem{Guillon02}
P.~Guillon, T.~Furon, and P.~Duhamel,
\newblock ``Applied public-key steganography,''
\newblock in {\em Proc. SPIE}, San Jose, CA, 2002.

\bibitem{Costa83}
M.~H.~M. Costa,
\newblock ``Writing on dirty paper,''
\newblock {\em IEEE Trans. Info. Thy}, vol. 29, no. 3, pp. 439--441, May 1983.

\bibitem{Cohen02DP}
A.~S. Cohen and A.~Lapidoth,
\newblock ``Generalized writing on dirty paper,''
\newblock in {\em Proc. Int. Symp. on Information Theory}, Jul. 2002.

\bibitem{Eggers02}
J.~J. Eggers, R.~Bäuml, R.~Tzschoppe, and B.~Girod,
\newblock ``Scalar {Costa} scheme for information embedding,''
\newblock {\em IEEE Trans. Signal Proc.}, 2002.

\bibitem{Anderson98}
R.~J. Anderson and F.~A.~P. Petitcolas,
\newblock ``On the limits of steganography,''
\newblock {\em IEEE Journal of Selected Areas in Communications}, vol. 16, no.
  4, pp. 474--481, 1998.

\bibitem{Miller04}
M.~L. Miller, G.~J. Doërr, and I.~J. Cox,
\newblock ``Applying informed coding and informed embedding to design a robust,
  high capacity watermark,''
\newblock {\em IEEE Trans. Image Proc.}, vol. 6, no. 13, pp. 792--807, 2004.

\bibitem{Chapel03}
V.~Chappelier, C.~Guillemot, and S.~Marinkovic,
\newblock ``Turbo trellis coded quantization,''
\newblock in {\em Proc. Int. Symposium on Turbo Codes}, Brest, France, Sep.
  2003.

\bibitem{Cachin98}
C.~Cachin,
\newblock ``An information-theoretic model for steganography,''
\newblock in {\em Proc. Int. Workshop on Information Hiding}, 1998.

\end{thebibliography}

\end{document}